\documentclass[review]{elsarticle}

\usepackage{lineno,hyperref}
\usepackage{graphicx}
\usepackage{amssymb}
\usepackage{amsmath}
\usepackage{epstopdf}
\usepackage{bm}
\modulolinenumbers[5]

\journal{Journal of \LaTeX\ Templates}

\begin{document}

\begin{frontmatter}

\title{Stability enhancement by induced synchronization using transient uncoupling in certain coupled chaotic systems}

\author[rvt]{G. Sivaganesh}
\author[focal]{A. Arulgnanam \corref{cor1}}
\ead{gospelin@gmail.com}
\author[focal1]{A. N. Seethalakshmi}
\cortext[cor1]{Corresponding author}
\address[rvt]{Department of Physics, Alagappa Chettiar Government College of Engineering $\&$ Technology, Karaikudi, Tamilnadu-630 004, India}
\address[focal]{Department of Physics, St. John's College, Palayamkottai, Tamilnadu-627 002, India, Affiliated to Manonmaniam Sundaranar University, Abishekapatti, Tirunelveli, Tamilnadu - 627 012, India}
\address[focal1]{Department of Physics, The M.D.T Hindu College, Tirunelveli, Tamilnadu - 627 010, India, Affiliated to Manonmaniam Sundaranar University, Abishekapatti, Tirunelveli, Tamilnadu - 627 012, India}


%
%

\begin{abstract}
In this work, we report the enhanced stability of induced synchronization observed through transient uncoupling in a class of unidirectionally coupled identical chaotic systems. The phenomenon of transient uncoupling implies the clipping of the chaotic attractor of the driven system in a drive-driven scenario and making the coupling strength active over the clipped regions. The {\emph{Master Stability Function}} (MSF) is used to determine the stability of the synchronized states for a finite clipping fraction in unidirectionally coupled chaotic systems subjected to transient uncoupling for fixed values of coupling strength. The effectiveness of transient uncoupling is observed through the existence of negative regions in the MSF spectrum for larger values of coupling strength. Further the two-parameter bifurcation diagram indicating the regions of stable synchronization for different values of clipping fraction and coupling strength has been obtained. The effect of the symmetry of chaotic attractors in enhancing the stability of synchronized states of coupled chaotic systems subjected to transient uncoupling is studied.
\end{abstract}


\end{frontmatter}

\linenumbers

\section{INTRODUCTION}

Synchronization is the dynamical process observed in weakly coupled chaotic systems \cite{Pecora1990,Pikovsky2003}. The phenomenon of chaos synchronization has been studied extensively in a variety of dynamical systems for a better understanding on its emergence and existence \cite{Ogorzalek1993,Rulkov1995,Rulkov1996,Rosenblum1996,Boccaletti2002}. The synchronization of two identical chaotic systems paves way for the transmission of signals and the coupled systems must necessarily exist in stable synchronized states for greater values of coupling strength which is most useful for practical applications of secure transformation of signals. Further, a greater symmetry (in size) of the chaotic attractor can effectively mask a large input signal in circuits \cite{Huang2005,Wang2016}. Obtaining a greater symmetry of the attractor through an increase in the break-point of the $voltage-current~(v-i)$ characteristics of the nonlinear element paves way to impose larger clipping on the phase-space of the attractors and thereby enhances the possibility of stable synchronization. The existence of similar dynamics of the drive and driven systems systems in the synchronization manifold has been indicated by the negative values of the  {\emph{Master Stability Function}} (MSF) \cite{Pecora1998}. Hence, the MSF of the coupled systems has been expected to be in the negative value regions for higher coupling strengths. The method of {\emph{transient uncoupling induced synchronization}} was introduced by  Schroder et al. \cite{Schroder2015} and proved to be an efficient method to enhance the stability of stable synchronized states in unidirectionally coupled chaotic systems \cite{Ghosh2018}. Enhancing the stability of synchronized states in coupled chaotic systems becomes necessary in understanding the collective dynamics of these systems in networks \cite{Liang2009}. However, only very few research work has been published in this thrust research area. In this paper we present the enhanced stable induced synchronization states observed through transient uncoupling in some prominent chaotic systems. This paper has been divided into two sections. In Section \ref{sec:2} the method of transient uncoupling is discussed in brief and in Section \ref{sec:3}, the enhancement of synchronization observed in coupled chaotic systems with attractors of different symmetry has been presented.

\section{Transient Uncoupling}
\label{sec:2}

In this section, we present the fundamental equations related to the {\emph{Master Stability Function}} (MSF) and the method of {\emph{transient uncoupling}}. The systems under consideration are unidirectionally coupled and hence the dynamics of the drive system is not affected with the change in the coupling parameter. When the drive and response systems are uncoupled, the isolated drive system is described by
\begin{equation}
\bf{\dot x} =   \bf{F(x)}
\label{eqn:1}
\end{equation}
where, {\bf{x}} is a $d$-dimensional vector and {\bf{F(x)}} is the velocity field. On introducing the concept of transient uncoupling, the response/driven system can be written as
\begin{equation}
\bf{\dot x_2} =   \bf{F(x_2)} + \epsilon \chi_{A} {\bf{G}}\otimes {\bf{E}}
\label{eqn:2}
\end{equation}
where {\bf{G}} is an  $N \times N$  matrix of coupling coefficients, $\bf{E}$ is an $n \times n$ matrix containing the information of the variables coupled and $\chi_A$ represents the transient uncoupling factor given as
\begin{equation}
\chi_A =
\begin{cases}
1 & \text{if ${\bf{x_2}} \in A$}\\
0 & \text{if ${\bf{x_2}} \notin A$}
\end{cases}
\label{eqn:3}
\end{equation}
where, $A$ is a region of phase space of the response system such that $A \subseteq \mathbb{R}^d$ in which the response system is controlled by the drive and the coupling is active. Hence, the coupling between the two systems is active only when $A$ is a subset of the phase space $\mathbb{R}^d$ of the response system and the coupling becomes normal for $A=\mathbb{R}^d$. The subset $A$ has been obtained by clipping a fraction of the phase space of the response system along a particular direction of the state variables given as
\begin{equation}
A_{\Delta} = \{ {\bf{x}}_2 \in \mathbb{R}^d : |{\bf{x}}_2 - {\bf{x}}_{2}^*| \le \Delta \}
\label{eqn:4}
\end{equation}
where ${\bf{x_2}}^{*}$ is a suitable point considered along the coordinate axis of the state variable $\bf{x}$. The variational equation of Eq. (2) is given by,
\begin{equation}
\dot \xi = [{\bf{I}}_N \otimes D{\bf{F}} + \epsilon \chi_{A} (\bf{G} \otimes  \bf{E})] \xi
\label{eqn:5}
\end{equation}
where $\bf{I}_N$ is an $N \times N$ identity matrix, D{\bf{F}} is the Jacobian of the uncoupled system and $\otimes$ represents the {\emph{inner}} or {\emph{Kronecker}} product. Hence, the coupling between the state variables is effective only on the existence of the response system within the clipped region. On diagonalization of the matrix {\bf{G}}, Eq. \ref{eqn:5} can be written as,
\begin{equation}
\dot {\xi_k} = [D{\bf{F}} + \delta \gamma_k \bf{E})] \xi_k,
\label{eqn:6}
\end{equation}
where $\delta = \epsilon \chi_{A}$ and $\gamma_k$ are the eigenvalues of {\bf{G}} with {\emph{k}} = 0 or 1. In general, the quantity $\delta \gamma_k$ are generally complex numbers which can be written in the form $\delta \gamma_k = \alpha + i \beta$. Hence, the general dynamical sytem is
\begin{equation}
\dot {\xi_k} = [D{\bf{F}} + (\alpha + i \beta) \bf{E})] \xi_k,
\label{eqn:7}
\end{equation}
The largest lyapunov exponent $\lambda^{\perp}_{max}$ of the generic variational equation in the transverse direction given by Eq. \ref{eqn:7} depending on $\alpha$ and $\beta$ values, is the {\emph{master stability function}} \cite{Pecora1998,Pecora1997}. 
The typical forms of the coupling matrix $\bf G$ and $\bf E$ are given as
\begin{equation*}
\bf{G} =
\begin{pmatrix}
-1 &&& 1 \\
 0 &&&  0 \\
\end{pmatrix},~
\bf{E} =
\begin{pmatrix}
1 &&& 0 &&& 0\\
0 &&& 0 &&& 0 \\
0 &&& 0 &&& 0\\
\end{pmatrix}. 
\end{equation*} 

\section{Results and Discussion}
\label{sec:3}
In this section we present the synchronization dynamics observed through transient uncoupling in different chaotic systems. The unidirectionally coupled {\emph{Chua's circuit}} systems is studied for the synchronization of identical chaotic attractors having two different symmetry followed by a study of the coupled {\emph{Rossler}} systems.

\subsection{Chua's circuit system}
\label{sec:2.1}

\begin{figure}
\begin{center}
\centering
\includegraphics[width=1\textwidth]{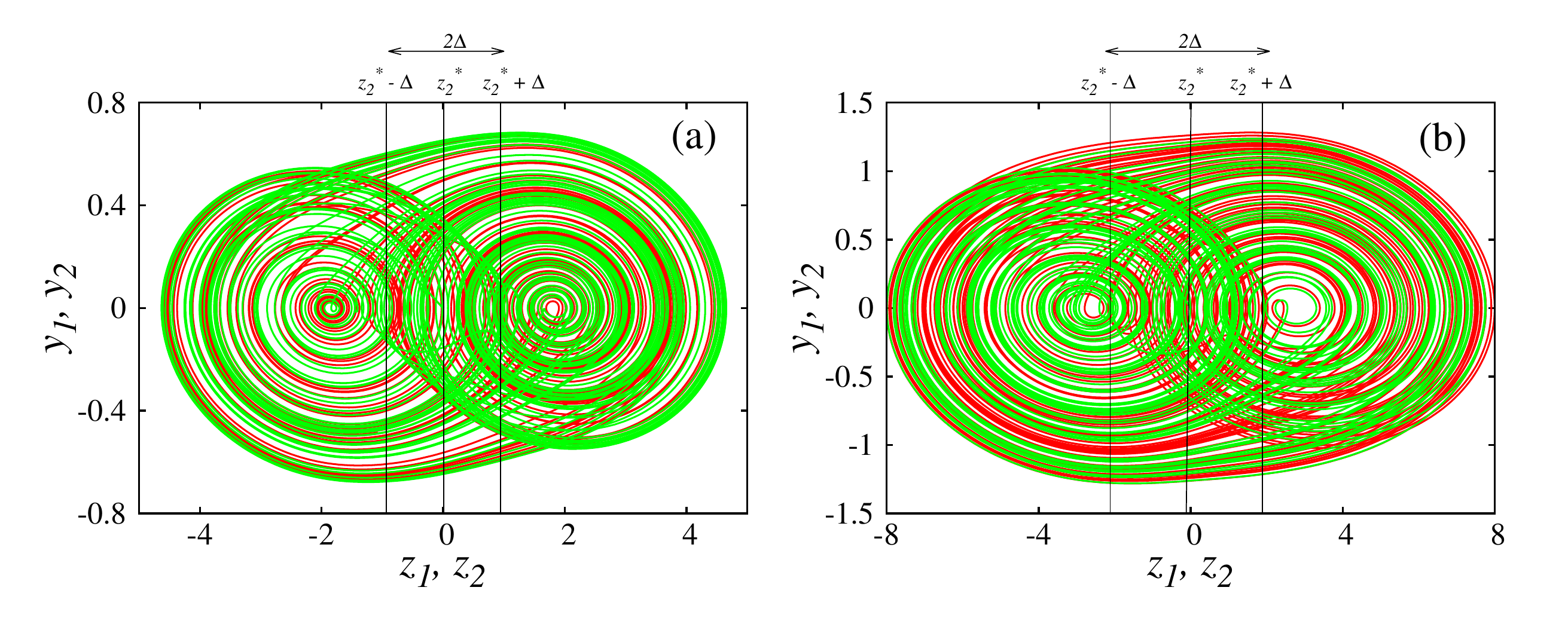}
\caption{(Color Online) Chaotic attractors of the drive (red) and driven systems (green) in the $z_{1}-y_{1}$ and $z_{2}-y_{2}$ planes for the {\emph{Chua's circuit}} system under the uncoupled state $(\epsilon = 0)$ with clipping of phase space of the driven system through transient uncoupling. (a) Chaotic attractors obtained for the system parameters $\alpha = 10, \beta = 14.87, \gamma=0, a=-1.27$ and $b=-0.68$; (b) Chaotic attractors obtained for the system parameters $\alpha = 10, \beta = 14.87, \gamma=0, a=-1.55$ and $b=-0.68$; }
\label{fig:1}
\end{center}
\end{figure}

In this section we discuss the effect of transient uncoupling induced synchronization observed in a third-order, autonomous system namely, the {\emph{Chua's circuit}} system. The {\emph{Chua's circuit}} has been identified as the first electronic circuit system to exhibit chaotic behavior in its dynamics \cite{Matsumoto1984,Matsumoto1985,Chua1992}. The normalized state equations of the unidirectionally coupled system under coupling of the $z$-variables subjected to transient uncoupling can be written as 
\begin{subequations}
\begin{eqnarray}
\dot x_1  &=&  \alpha (y_1-x_1+f(x_1)), \\ 
\dot y_1  &=&  x_1-y_1+z_1,\\ 
\dot z_1  &=&  -\beta y_1 - \gamma z_1,\\
\dot x_2  &=&  \alpha (y_2-x_2+f(x_2)), \\ 
\dot y_2  &=&  x_2-y_2+z_2,\\ 
\dot z_2  &=&  -\beta y_2 - \gamma z_2+ \epsilon \chi_A (z_1 - z_2),
\end{eqnarray}
\end{subequations}
where $f(x_{1}), f(x_2)$ represent the three-segmented piecewise-linear function of the drive and driven systems given as 
\begin{equation}
f(x_{1,2}) =
\begin{cases}
-bx_{1,2}+(a-b) & \text{if $x_{1,2} > 1$}\\
-ax_{1,2} & \text{if $|x_{1,2}| < 1$}\\
-bx_{1,2}+(a-b) & \text{if $x_{1,2} < -1$}
\end{cases}
\label{eqn:16}
\end{equation}
where the state variables $x_1,y_1,z_1$ and $x_2,y_2,z_2$ corresponds to the drive and the driven units, respectively. The driven system has been coupled to the drive through the
$z$-variable and by the transient uncoupling factor $\epsilon \chi_A$. The normalized parameters of the system are fixed at $\alpha = 10, \beta = 14.87, \gamma=0$ and $b=-0.68$. The parameter $a$ has been kept at two different values $a=-1.27$ and $a=-1.55$ to obtain chaotic attractors of different symmetry. The {\emph{Chua's}} circuit exhibits a period-doubling route to chaos. The double-scroll chaotic attractor of the {\emph{Chua's circuit}} systems corresponding to the drive (red) and the driven (green) systems in the $z_{1}-y_{1}$ and $z_{2}-y_{2}$ phase planes under the uncoupled state $(\epsilon=0)$ with clipping of the phase space of the driven system along the $z_2$-coordinate axis for the parameter $a=-1.27$ and $a=-1.55$ are shown in Fig. \ref{fig:1}(a) and \ref{fig:1}(b), respectively. The term $2 \Delta$ represents the clipped region along the $z$-axis over which the coupling strength is active. The clipping of the phase space of chaotic attractors of the driven system by transient uncoupling over which the coupling between the drive and driven systems is active has been represented by the $2\Delta$ region in Fig. \ref{fig:1}. The point ${z_2}^*$ has been considered as the center of the chaotic attractor and the clipping of phase space has been performed to a width of $\Delta$ along the coordinate axis ($z$-axis) on either side of ${z_2}^*$, resulting in a clipping width of $2\Delta$.

\begin{figure}
\begin{center}
\centering
\includegraphics[width=1\textwidth]{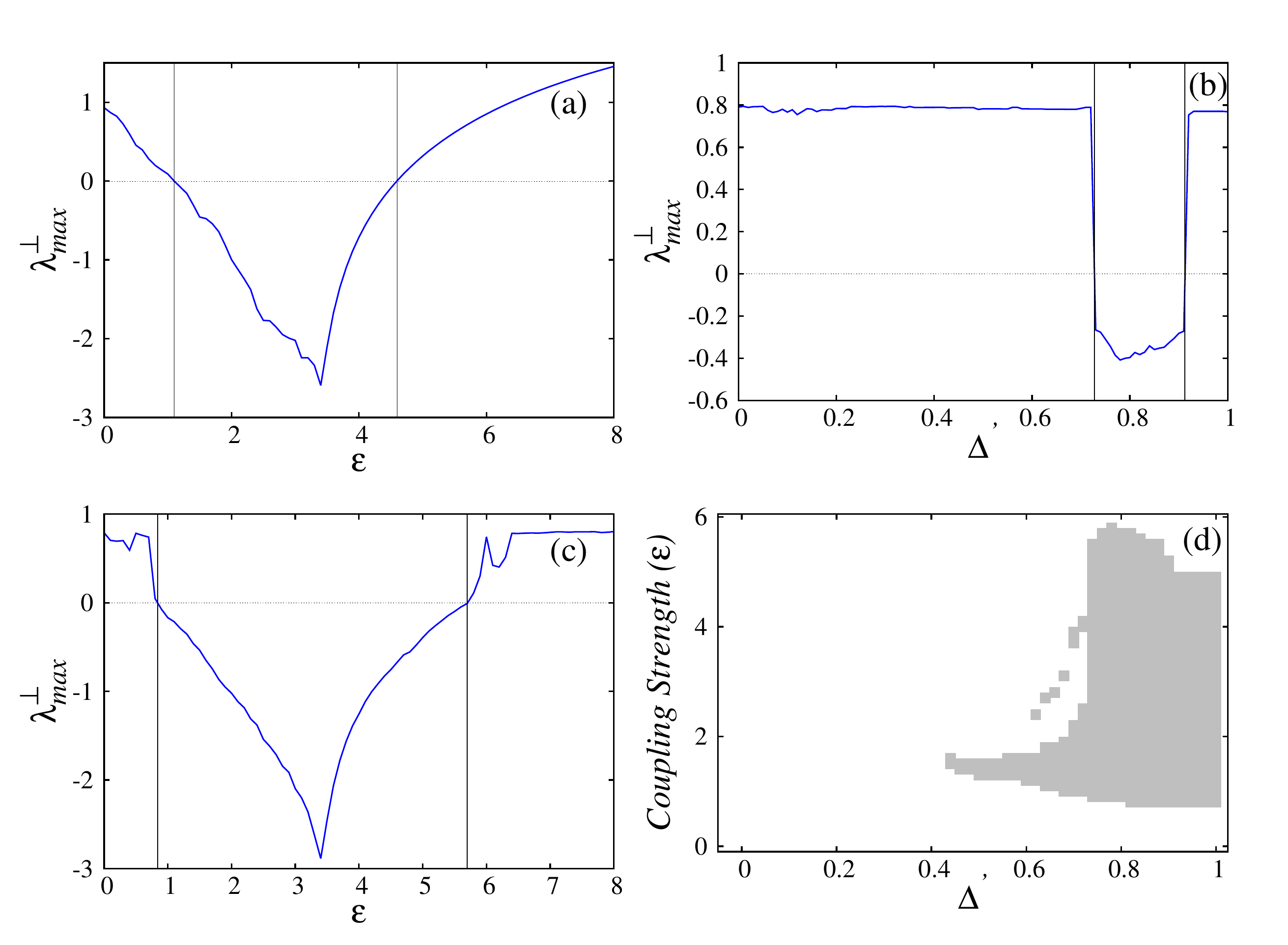}
\caption{(Color Online) Stability of synchronization in {\emph{Chua's circuit}} systems for the parameters $\alpha = 10, \beta = 14.87, \gamma=0, a=-1.27$ and $b=-0.68$. (a) MSF of the {\emph{Chua's circuit}} system obtained without transient uncoupling. The coupled system exists in stable synchronized states in the region $1.1 \le \epsilon \le 4.59$; (b) MSF as a function of the clipping fraction $\Delta^{'}$ for $\epsilon = 5$ indicating the stable synchronized states for higher clipping fractions and unsynchronized states for lower clipping fractions; (c) MSF as a function of the coupling parameter under transient uncoupling obtained with ${z_2}^{*} = 0.01$ and $\Delta^{'} = 0.8$. Enhancement of stable synchronized states represented by the larger negative valued regions of MSF in the range $0.84 \le \epsilon \le 5.71$; (d) Two-parameter bifurcation diagram in the $\Delta^{'}-\epsilon$ plane indicating the parameter regions for stable synchronized states (gray color).}
\label{fig:2}
\end{center}
\end{figure}
\begin{figure}
\begin{center}
\centering
\includegraphics[width=1\textwidth]{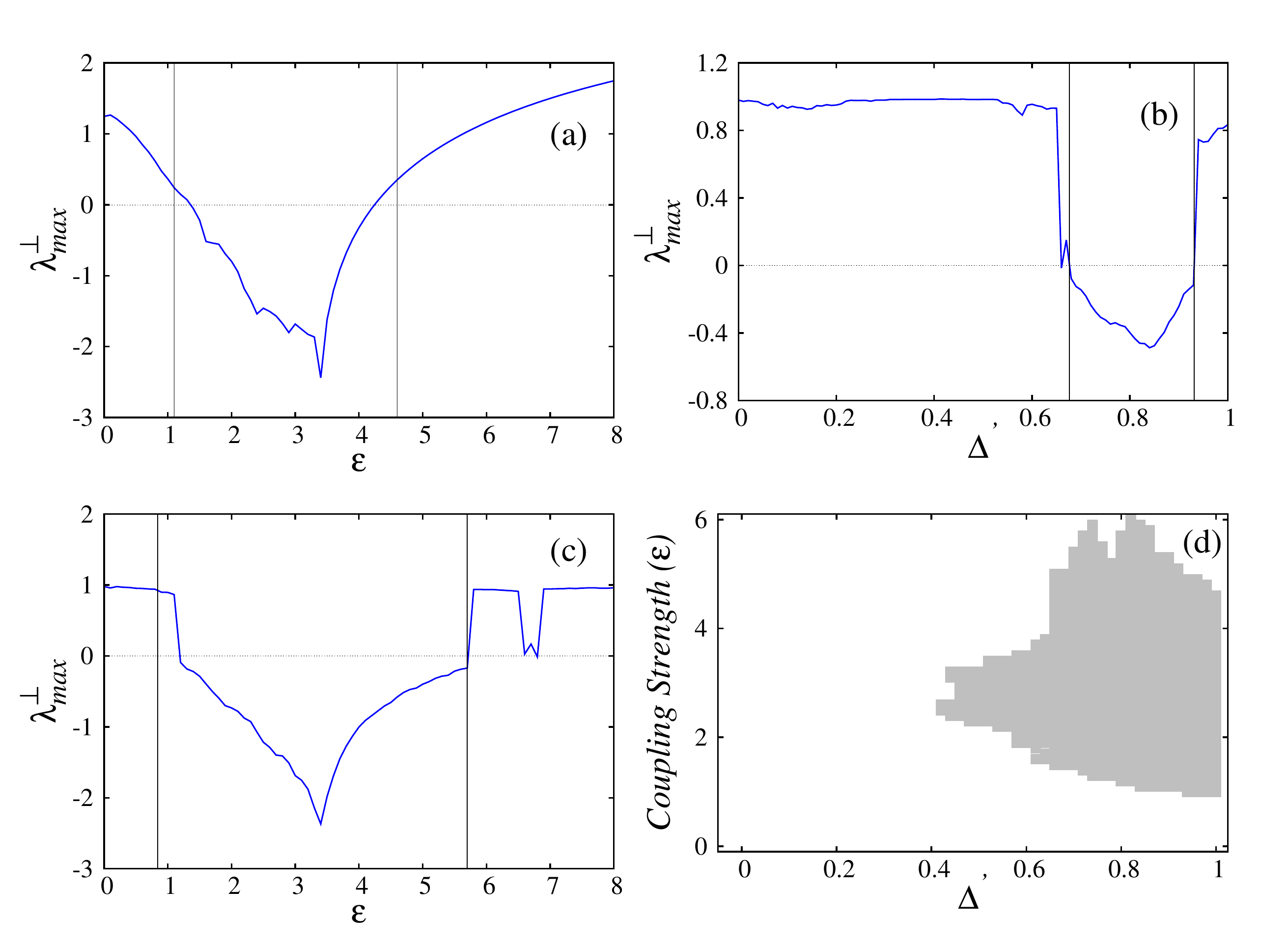}
\caption{(Color Online) Stability of synchronization in coupled {\emph{Chua's circuit}} systems for the parameters $\alpha = 10, \beta = 14.87, \gamma=0, a=-1.55$ and $b=-0.68$. (a) MSF of the {\emph{Chua's circuit}} system obtained without transient uncoupling. The coupled system exists in stable synchronized states in the region $1.36 \le \epsilon \le 4.25$; (b) MSF as a function of the clipping fraction $\Delta^{'}$ for $\epsilon = 5$ indicating the stable synchronized states for higher clipping fractions and unsynchronized states for lower clipping fractions; (c) MSF as a function of the coupling parameter under transient uncoupling obtained with ${z_2}^{*} = -0.116$ and $\Delta^{'} = 0.8$. Enhancement of stable synchronized states represented by the larger negative valued regions of MSF in the range $1.19 \le \epsilon \le 5.715$; (d) Two-parameter bifurcation diagram in the $\Delta^{'}-\epsilon$ plane indicating the parameter regions for stable synchronized states (gray color).}
\label{fig:3}
\end{center}
\end{figure}

The synchronization dynamics of the system under transient uncoupling for the parameters $\alpha = 10, \beta = 14.87, \gamma=0, a=-1.27$ and $b=-0.68$ is shown in Fig. \ref{fig:2}. In the absence of transient uncoupling the variation of the MSF as a function of the coupling parameter shows stable synchronized states in the range $1.1 \le \epsilon \le 4.59$. Hence, the coupled system becomes unsynchronized for coupling strengths $\epsilon > 4.59$ under normal coupling. Now we introduce the transient uncoupling factor in the system equations and study its synchronization stability. The stability of the synchronized states can be analyzed for different regions of clipping which can be expressed by the clipping fraction $\Delta^{'} = 2\Delta/\Omega$ where, $\Omega$ is the width of the attractor along the clipping coordinate axis (z-axis). The variation of the MSF of the coupled system  under transient uncoupling as a function of the clipping fraction with the coupling strength fixed at $\epsilon = 5$ is shown in Fig. \ref{fig:2}(b). For $\Delta=0$, the two systems are uncoupled and hence does not synchronize and for $\Delta = \Omega/2$, the coupling between the systems become normal and the original unsynchronized behavior of the systems under normal coupling is obtained. From Fig. \ref{fig:2}(b) it is observed that the coupled systems enter into stable synchronized states only for larger values of the clipping fraction $\Delta^{'}$. Figure \ref{fig:2}(c) showing the variation of the MSF with the coupling strength indicates a broader stable synchronized state in the range $0.84 \le \epsilon \le 5.71$ for a clipping fraction of $\Delta^{'}=0.8$. The introduction of transient uncoupling induces synchronization in the higher coupling strength regions for which synchronization behavior has not been observed under normal coupling. The two-parameter bifurcation diagram obtained in the $\Delta^{'}-\epsilon$ plane showing the parameter regions for the existence of the coupled system in the synchronized state is as shown in Fig. \ref{fig:2}(d). The gray colored regions observed at larger clipping fractions gives the parameters $\Delta^{'}$ and $\epsilon$ for which the MSF is negative. Hence, the {\emph{Chua's circuit}} exhibits stable synchronized states for larger coupling strengths at higher values of clipping fraction.\\
The synchronization dynamics of coupled {\emph{Chua's circuit}} systems obtained for the parameters $\alpha = 10, \beta = 14.87, \gamma=0, a=-1.55$ and $b=-0.68$ is presented in Fig. \ref{fig:3}. The chaotic attractor obtained for these parameters presents a greater symmetry as shown in Fig. \ref{fig:1}(b). Under normal coupling, the coupled systems exhibit stable synchronized states in the range of the coupling strengths $1.36 \le \epsilon \le 4.25$. Figure \ref{fig:3}(b) showing the variation of the MSF with clipping fraction for the coupling strength fixed at $\epsilon=5$ indicates a broader stable synchronized region as compared to Fig. \ref{fig:2}(b). Hence, it is evident that a greater symmetry of the chaotic attractor promises synchronization at comparably lower clipping fractions, provided the coupling strength remains the same. The variation of MSF with coupling strength showing a broad stable synchronized region in the range $1.19 \le \epsilon \le 5.715$ for a clipping fraction $\Delta^{'}=0.8$ is shown in Fig. \ref{fig:3}(c). Figure \ref{fig:3}(d) shows the two-parameter bifurcation diagram obtained in the $\Delta^{'}-\epsilon$ plane giving the parameter region for the stable synchronized states. The gray colored regions shown in Fig. \ref{fig:3}(d) obtained for the chaotic attractor of a greater symmetry spreads over a broader range of clipping fraction as compared to that of a lesser symmetry one shown in Fig. \ref{fig:2}(d). Hence, a greater symmetry of the chaotic attractor enhances stable synchronization of the coupled systems over a broader region of clipping fraction.\\

\subsection{Rossler system}
\label{sec:2.2}

\begin{figure}
\begin{center}
\includegraphics[width=1\textwidth]{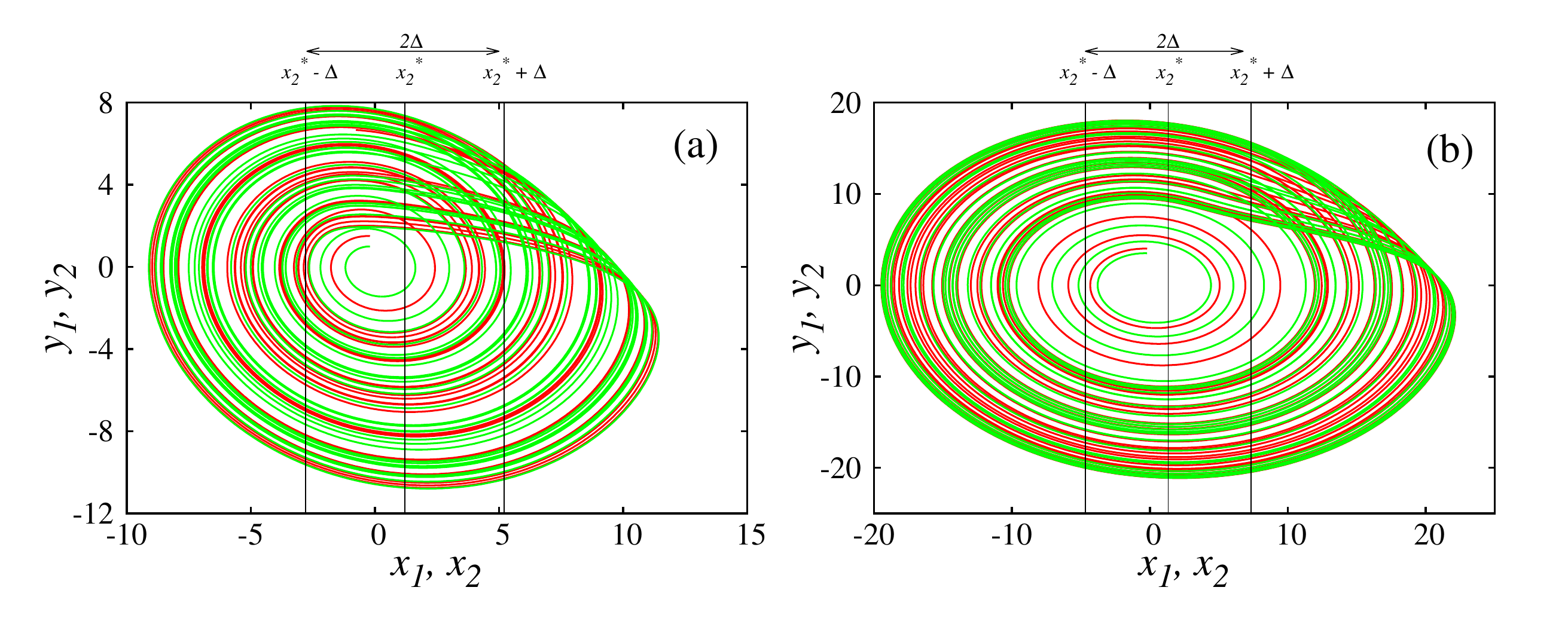}
\caption{(Color Online) Chaotic attractors of the drive (red) and response systems (green) in the $x_{1}-y_{1}$ and $x_{2}-y_{2}$ planes for the {\emph{Rossler}} system under uncoupled state $(\epsilon = 0)$ with clipping of phase space of the response system through transient uncoupling. The coupling strength $\epsilon$ is active only within the clipped region $2\Delta$.  (a) Chaotic attractor obtained for the parameters $a=0.2,b=0.2,c=5.7$; (b) Chaotic attractor obtained for the parameters $a=0.1,b=0.1,c=14$.}
\label{fig:5}
\end{center}
\end{figure}

\begin{figure}
\begin{center}
\includegraphics[width=1\textwidth]{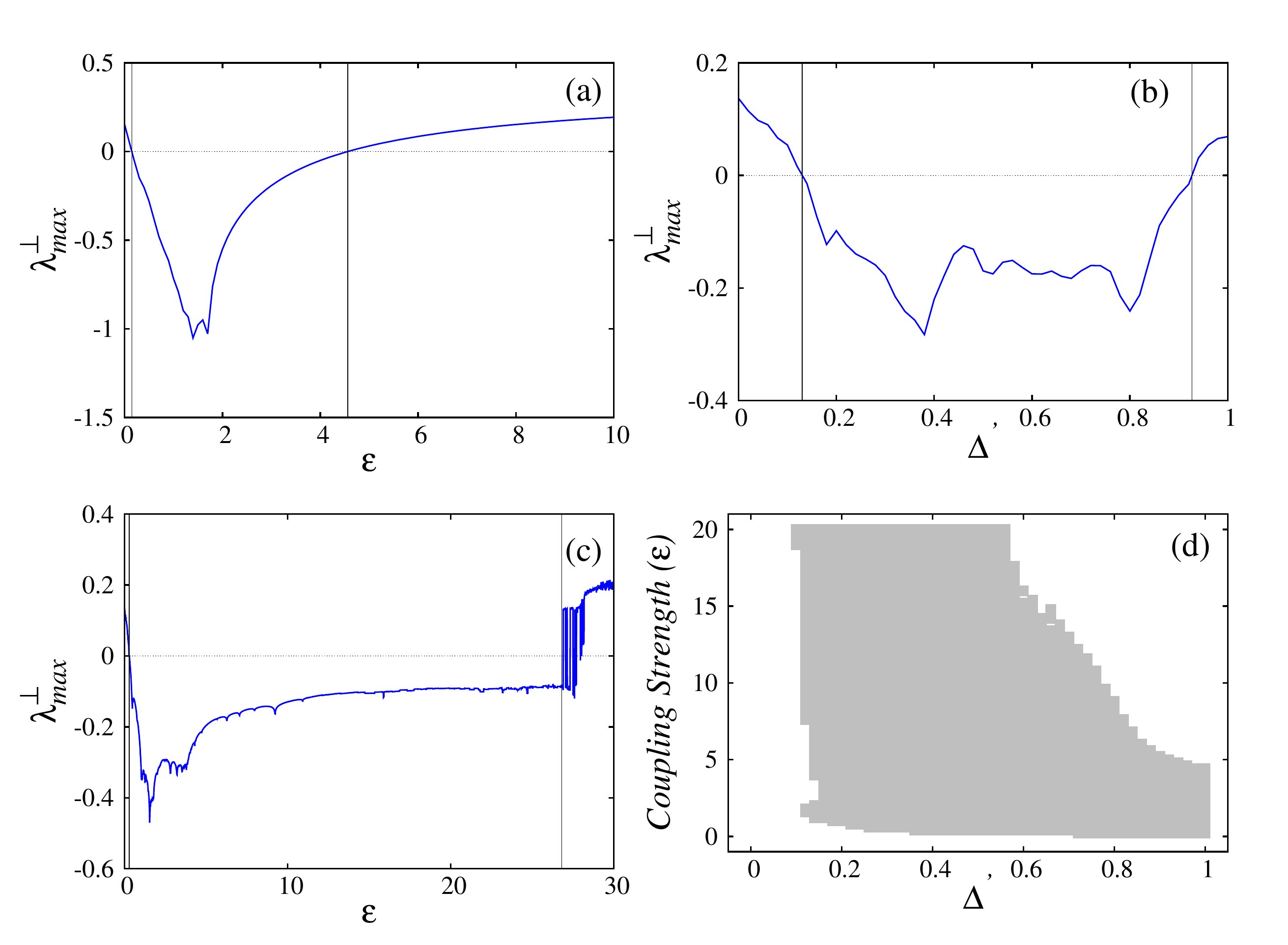}
\caption{(Color Online) Stability of synchronization in coupled {\emph{Rossler}} systems for the parameters $a=0.2,b=0.2,c=5.7$. (a) MSF of the Rossler system obtained without transient uncoupling. The coupled system exists in stable synchronized states in the region $0.149 \le \epsilon \le 4.56$; (b) MSF as a function of the clipping fraction $\Delta^{'}$ for $\epsilon = 5$ indicating the stable synchronized states for moderate clipping and unsynchronized states for lower and higher clipping fractions; (c) MSF as a function of the coupling parameter under transient uncoupling obtained with ${x_2}^{*} = 1.2$ and $\Delta^{'} = 0.25$. Enhancement of stable synchronized states represented by the larger negative valued regions of MSF in the range $0.286 \le \epsilon \le 26.8$; (d) Two-parameter bifurcation diagram in the $\Delta^{'}-\epsilon$ plane indicating the parameter regions for stable synchronized states (gray color).}
\label{fig:6}
\end{center}
\end{figure}

The dynamical equations of unidirectionally coupled {\emph{Rossler}} systems \cite{Rossler1976} subjected to transient uncoupling can be written as
\begin{subequations}
\begin{eqnarray}
\dot x_1  &=&  y_1 - z_1, \\ 
\dot y_1  &=&  x_1 + a y_1,\\ 
\dot z_1  &=&  b + (x_1 - c) z_1,\\
\dot x_2  &=&  y_2 - z_2 + \epsilon \chi_A (x_1 - x_2), \\ 
\dot y_2  &=&  x_2 + a y_2,\\ 
\dot z_2  &=&  b + (x_2 - c) z_2,
\end{eqnarray}
\end{subequations}
where the state variables $x_1,y_1,z_1$ and $x_2,y_2,z_2$ corresponds to the drive and the driven units, respectively. The driven system is coupled to the drive through the $x$-variable and by the transient uncoupling factor $\epsilon \chi_A$. The chaotic attractors of the drive (red) and the driven (green) {\emph{Rossler}} systems in the $x_1-y_1$ and $x_2-y_2$ phase planes in the uncoupled state ($\epsilon=0$) is as shown in Fig. \ref{fig:5}(a) and \ref{fig:5}(b) for the parameters $a=0.2,b=0.2,c=5.7$ and $a=0.1, b=0.1, c=14$, respectively. The point ${x_2}^*$ has been considered as the center of the chaotic attractor and the clipping of phase space has been performed to a width of $\Delta$ along the $x$-axis on either side of ${x_2}^*$ with a clipping width of $2\Delta$. The MSF of the coupled {\emph{Rossler}} systems under normal coupling is as shown in Fig. \ref{fig:6}(a). The lower and upper bound for stable synchronization shown in Fig. \ref{fig:6}(a) is given as $0.149 \le \epsilon \le 4.56$. The coupled system loses its synchronization stability for the coupling strengths $\epsilon > 4.56$.
\begin{figure}
\begin{center}
\includegraphics[width=1\textwidth]{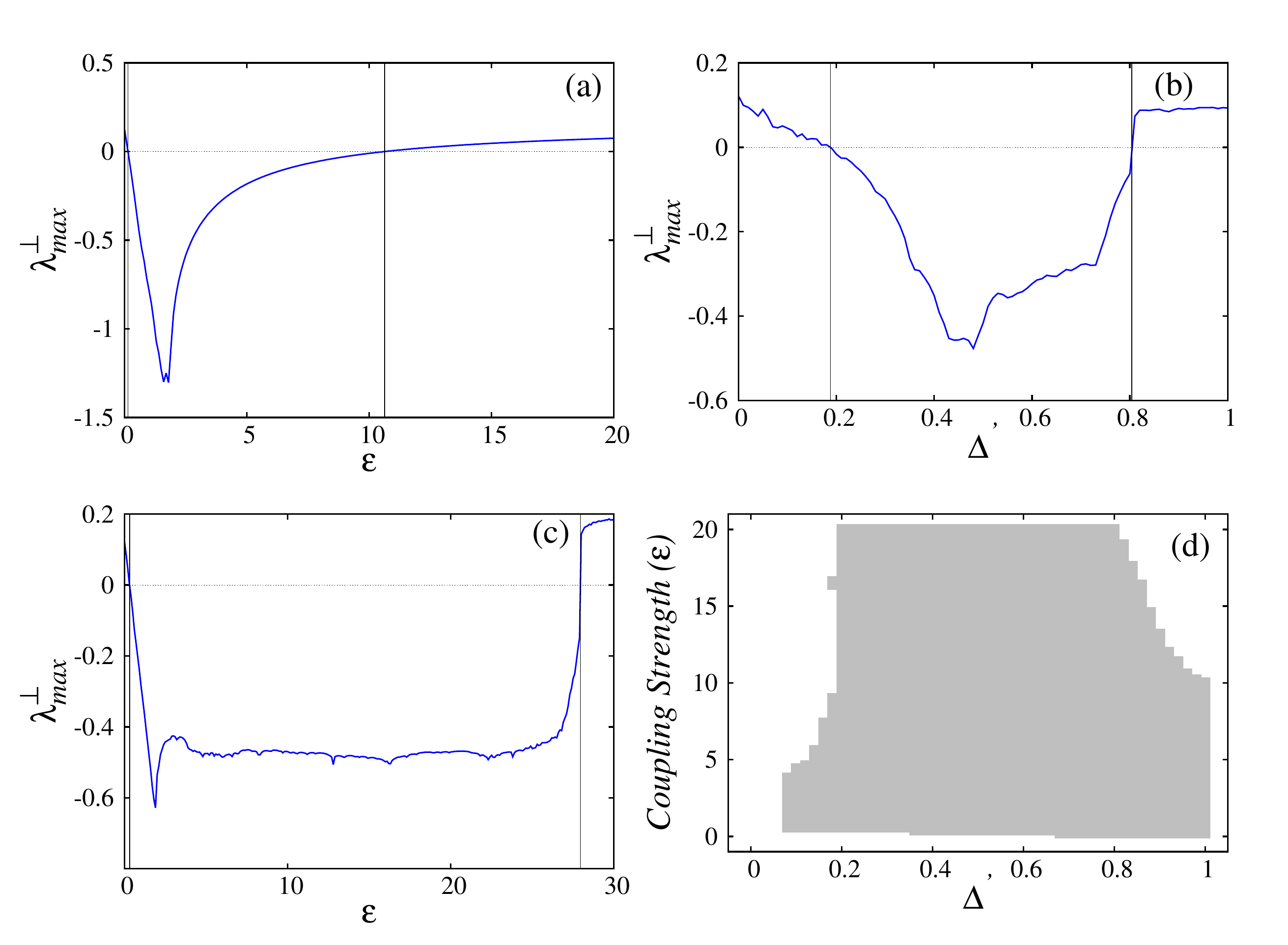}
\caption{(Color Online) Stability of synchronization in coupled {\emph{Rossler}} systems for the parameters $a=0.1,b=0.1,c=14$. (a) MSF of the Rossler system obtained without transient uncoupling. The coupled system exists in stable synchronized states in the region $0.142 \le \epsilon \le 10.63$; (b) MSF as a function of the clipping fraction $\Delta^{'}$ for $\epsilon = 20$ indicating the stable synchronized states for moderate clipping and unsynchronized states for lower and higher clipping fractions; (c) MSF as a function of the coupling parameter under transient uncoupling obtained with ${x_2}^{*} = 1.325$ and $\Delta^{'} = 0.25$. Enhancement of stable synchronized states represented by the larger negative valued regions of MSF in the range $0.307 \le \epsilon \le 27.95$; (d) Two-parameter bifurcation diagram in the $\Delta^{'}-\epsilon$ parameter plane indicating the parameter regions for stable synchronized states (gray color).}
\label{fig:7}
\end{center}
\end{figure}
The MSF of the coupled system under transient uncoupling as a function of the clipping fraction with the coupling strength fixed at $\epsilon = 5.0$ is as shown in Fig. \ref{fig:6}(b). From Fig. \ref{fig:6}(b) it is evident that for moderate values of clipping fraction the coupled systems exist in stable synchronized states even for the coupling strength $\epsilon=5.0$. Figure \ref{fig:6}(c) shows the variation of the MSF as a function of the coupling strength for the parameters ${x_2}^{*} = 1.2$ and $\Delta^{'} = 0.25$. Hence, for a moderate clipping fraction of $\Delta^{'} = 0.25$, the coupled system shows an enhanced stable synchronized state in the range $0.286 \le \epsilon \le 26.8$. The two-parameter bifurcation diagram shown in Fig. \ref{fig:6}(d) obtained in the $\Delta^{'}-\epsilon$ plane gives the parameter regions for which the coupled systems exist in the stable synchronized state. The gray colored regions gives the parameters $\Delta^{'}$ and $\epsilon$ for which the MSF is negative.\\
The stability of coupled {\emph{Rossler}} systems existing in the chaotic state shown in Fig. \ref{fig:5}(b) obtained for the system parameters $a=0.1,b=0.1,c=14$ has been studied in a similar way explained above. The MSF of the coupled system under normal coupling shown in Fig. \ref{fig:7}(a) indicates the existence of the system in stable synchronized states for the values of coupling strength in the range $0.142 \le \epsilon \le 10.63$. Figure \ref{fig:7}(b) showing the variation of the MSF with the clipping fraction $\Delta^{'}$ for a fixed value of the coupling strength $\epsilon = 20$ indicates greater stable synchronized states for moderate clipping fractions. The variation of the MSF  as a function of the coupling strength for the parameters ${x_2}^{*} = 1.325$ and $\Delta^{'} = 0.25$ is as shown in Fig. \ref{fig:7}(c). From Fig. \ref{fig:7}(c) it is observed that under moderate values of clipping the coupled systems exist in stable synchronized states for larger values of coupling strength. The two-parameter bifurcation diagram shown in Fig. \ref{fig:7}(d) gives the parameters $\Delta^{'}$ and $\epsilon$ for which the MSF is negative (gray colored region).\\

%
%

\section{Conclusion}

In this paper we have presented the stability of the induced synchronized states observed in unidirectionally coupled chaotic systems subjected to transient uncoupling. Two prominent chaotic systems namely, the {\emph{Chua's circuit}} and the {\emph{Rossler}} systems, are studied for the induced synchronization phenomenon obtained through transient uncoupling. The introduction of transient uncoupling greatly enhances the stability of the stable synchronized states. In particular, the upper bound for synchronization has been pushed to greater values of coupling strength marking the enhancement of stable synchronized states. Hence, the coupled systems persist in the synchronized state for greater values of coupling strength than that under normal or standard coupling. The proliferation of the negative eigenvalue regions of the state variable corresponding to the clipped direction has been identified as a factor inducing synchronization of the coupled systems. Further, it is observed that an attractor with a greater symmetry exhibits enhanced stable synchronized states over a larger range of clipping fractions than the attractor with a smaller symmetry. Hence, the symmetry of the chaotic attractor plays an important role in the enhancement of stable synchronized states under transient uncoupling for higher values of coupling strengths. The enhancement of stability of the synchronized states is of greater importance in studying the collective dynamics of these systems in large networks and also for the application of these systems for secure transmission of information.

\section*{References}

\bibliography{mybibfile}

\end{document}